\input amstex
\documentstyle{amsppt}
\NoBlackBoxes
\TagsOnRight
\CenteredTagsOnSplits

\magnification=1200
\hcorrection{-0.0625 in}
\vcorrection{0.0 in}
\pagewidth{5.6 in}
\pageheight{7.5 in}
\nopagenumbers

\topmatter

\title Graphical Representations for Ising Systems in External Fields
\endtitle

\leftheadtext { L. Chayes, J. Machta and O. Redner }
\rightheadtext\nofrills {Ising Clusters in External Fields}

\author
\hbox{\hsize=2.8in
\vtop{\centerline{L. Chayes}
\centerline{{\it Department of Mathematics}}
\centerline{{\it University of California,}}
\centerline{{\it Los Angeles, California, USA}}}
\vtop{\centerline{J. Machta}
\centerline{{\it Department of Physics and Astronomy}}
\centerline{{ \it University of Massachusetts, }}
\centerline{{ \it  Amherst, Massachusetts, USA}}}}{\rm and}
\\ O. Redner\\ {\it Institut f\"ur Theoretische Physik\\ Universit\"at
T\"ubingen, Germany}
\endauthor

\address L. Chayes
\hfill\newline Department of Mathematics
\hfill\newline University of California
\hfill\newline Los Angeles, California 90095-1555
\endaddress
\email lchayes\@math.ucla.edu
\endemail
\address J. Machta
\hfill\newline Department of Physics and Astronomy,
\hfill\newline University of Massachusetts,
\hfill\newline Amherst, MA 01003-3720
\endaddress
\email machta\@phast.umass.edu
\endemail

\address O. Redner
\hfill\newline Institut f\"ur Theoretische Physik,
\hfill\newline Universit\"at T\"ubingen,
\hfill\newline Auf der Morgenstelle 14,
\hfill\newline D-72076 T\"ubingen, Germany
\endaddress
\email oliver.redner\@student.uni-tuebingen.de
\endemail
\thanks JM was partially supported by the NSF under grant DMR-9632898,
LC was partially supported by the NSA under grant MDA904-98-1-0518
\endthanks
\abstract
\medskip

A graphical representation based on duplication is developed that is
suitable for the study of Ising systems in external fields. Two independent
replicas of the Ising system in the same field are treated as a single
four-state (Ashkin-Teller) model.  Bonds in the graphical representation
connect the Ashkin-Teller spins. For ferromagnetic systems it is proved
that ordering is characterized by percolation in this representation.  The
representation leads immediately to cluster algorithms; some applications
along these lines are discussed.
\endabstract
\endtopmatter

\document
\baselineskip = 12pt
\newpage
\subhead Introduction
\endsubhead Graphical representations have provided a powerful tool for
both the analytical and numerical study of statistical mechanics systems.
However, the full power of these methods can only be realized when the
phases and the phase transitions in the original system are easily
identifiable in the graphical representation.  Usually this involves the
{\it percolation} properties of the graphical representation.
Consequently, the subject is still something of an ``art''; various systems
are treated on a case--by--case basis.  Despite the recent progress on
systems other than Potts models (c.f. \cite {CM II} and references therein)
an apparent requirement is a high degree of internal symmetry for the
individual degrees of freedom.

In this paper, we will treat the simplest possible system with {\it no}
internal symmetry, namely the ferromagnetic Ising model in an external
field. The representation is actually that of a duplicated (or
``replicated'') system and is closely related to the so called red/blue
representation of the Ashkin--Teller model that was introduced in
\cite{CM I}.  Before launching into a full description of this
representation -- the subject of the next section -- let us illustrate the
problems that beset the standard approaches.

Consider the usual Ising Hamiltonian:
$$
\Cal H = -\sum_{\langle i,j \rangle}J_{i,j}\sigma_i\sigma_j - \sum_ih_i
\sigma_i
\tag 1
$$ where the bonds $(\langle i,j \rangle)$ and sites $(i)$ are part of some
regular graph, the $J_{i,j}$ are positive and $-\infty < h_i < + \infty$.
To simplify a number of technical points, it will be assumed that the model
is {\it short ranged}, that is to say, the number of bonds emanating from
any site is uniformly bounded.

 In case $h_i \equiv 0$, there is the standard random cluster representation
\cite {FK} for bond configurations $\omega$:  In finite volume let
$$ W(\omega) = B_{\bold J}(\omega)2^{c(\omega)}
\tag 2
$$ where $B_{\bold J}$ is the Bernoulli factor of $1 - e^{-2\beta J_{i,j}}$
for each occupied bond of $\omega$ and $e^{-2\beta J_{i,j}}$ for each
vacant bond and $c(\omega)$ is the number of components.  (Here we ignore
modifications due to boundary conditions.)  Since, for each $\omega$, these
quantities (weights) are non--negative they define a probability measure on
the collection of all bond configurations.  As such, these measures admit
infinite volume limits.  It is remarked that for regular
infinite volume systems, e\.g\. nearest neighbor models on
$\Bbb Z^d$, the percolation of these bonds corresponds to positive
spontaneous magnetization in the spin--system \cite {CK, ACCN}.

For $h_i$ not identically zero, there are two obvious approaches both of
which lead to unsatisfactory results:  First we may include the magnetic
fields directly in the formula for the weights.  The result is
$$ W(\omega) = B_{\bold J}(\omega)\prod_{\alpha = 1}^{c(\omega)} 2
\cosh[\bold h(K_\alpha)]
\tag 3
$$ where the product runs over the clusters $K_1, \dots K_{c(\omega)}$ and
$\bold h(K_\alpha) = \beta \sum_{i \in K_\alpha}h_i$.   In other words each
cluster is given a weight in accord with the field energy obtained if all
spins of the cluster are of the same sign.  The second approach -- which is
most useful if all the $(h_i)$ are of the same sign -- is the introduction
of ghost bonds.  In general, the ghost system is defined as follows:
clusters  are defined and preliminarily weighted as in the zero field
case.   In addition, for every site $i$ such that
$h_i > 0$, there may be an ``up'' ghost bond which carries the weight
$1 - e^{-2\beta h_i}$ or no such bond with weight $e^{-2\beta h_i}$.
Similarly, if $h_j < 0$, the site $j$ may get a ``down'' ghost bond with
weight
$1 - e^{-2\beta |h_j|}$, etc.  However, the system is constrained so that
no cluster may have ghost bonds of both types.  Thus, if $\overline \omega$
denotes the extended system of ordinary and ghost bonds the weight for
$\overline \omega$ is given by
$$
\overline W(\overline \omega) = \Phi(\overline \omega)B_{\bold J, \bold
h}(\overline
\omega)
\prod_{\alpha = 1}^{c(\omega)} v_\alpha;
\tag 4
$$
here $\Phi(\overline \omega)$ is the function that is one if the above
mentioned constraint is satisfied and zero otherwise, $v_\alpha$ is 2 if
there are no ghost bonds in the $\alpha^{\text{th}}$ cluster and 1
otherwise and $B_{\bold J, \bold h}(\overline \omega)$ is the Bernoulli
weight for the configuration $\overline \omega$:
$1-e^{2\beta J_{i,j}}$  ($e^{2\beta J_{i,j}}$) if the bond $\langle i,j
\rangle$ is occupied (unoccupied) and
$1-e^{2\beta |h_i|}$ ($e^{2\beta |h_i|}$) if the ghost bond at $i$ is
occupied (unoccupied).  It is not hard to show that the weights in Eqn\.
(3) are obtained from those in Eqn\. (4) by summing over the ghost bonds.

The difficulty with either  of these representations is manifestly
apparent.  Consider the case
$h_i \equiv h > 0$  on the usual $d$--dimensional lattice with $d \geq 2$.
By the Lee--Yang theorem, there is no phase transition as temperature is
varied.  Nevertheless, it is clear that in the graphical representation,
there is a percolation transition at some finite $\beta_f(h) \leq \beta_c$.
Thus, these representations are susceptible to false percolation
transitions.  In less trivial situations such as random field problems, it
can be argued that such false percolations will also occur.   Cluster
algorithms based on this representation are ineffective because of the
weighting of cluster flips and the problem of false percolation:
Notwithstanding the fact that large--scale clusters may be of no particular
interest, they  tend  to become dynamically frozen.  For example, in the
second representation, no cluster can flip unless it is devoid of ghost
bonds which  occurs very rarely for large clusters. These effects lead to
severe slowing down both for critical and non--critical values of the
parameters.

Attempts have been made to circumvent these difficulties by restricting the
size of clusters (e\.g\. \cite{NB}) -- with some degree of success -- but
in principal, there can be no percolative signal of the phase transition in
any such representation.

It is therefore of interest to produce representations for spin--systems in
arbitrary external fields; the subject of this work.   In this regard, we
have recently learned of a related approach to these sets of problems that
is applicable to systems with {\it periodically} varying fields.  These
methods are described in
\cite{HBb} and \cite{HBc} and were applied to the hard--core lattice gas (a
limiting case of the staggered field problem) \cite{HBa}, \cite{HBb} and
related systems
\cite{HBc}.  We shall describe the similarities (and differences) between
this method and the current method in the conclusions section. At present,
these approaches are, by and large, restricted to Ising--type systems.
Nevertheless, this covers a number of problems that are of current interest
such as the random--field Ising model. In the context of the
staggered--field Ising system, an algorithm based on this representation
has already been tested
\cite{RMC} with quite satisfactory results.
\subhead Red/Blue Representation in Zero Field
\endsubhead Let us start with the zero field case.  Much of this material
appears in
\cite{CM I} (Appendix B3) in the context of the Ashkin--Teller model.
Consider the duplicated system so that at each site $i$, we have
$\sigma_i = \pm 1$, $\tau_i = \pm 1$.  We will envision the duplicated
problem as a 4--state spin model, in particular,  the 4--state clock
model.  Let us use $s_i$ to denote the spin variable, regarded as one of
four points on the circle and $\underline s$ as notation for a spin
configuration. If $\langle i,j \rangle$ is a bond of the lattice, there are
three possible energy states for the spins $s_i$ and $s_j$, the highest of
which is set to zero without loss of generality. Thus we define
$\Cal E^0_{i,j} = 4J_{i,j}$ and $\Cal E^1_{i,j} = 2J_{i,j}$.   The
Hamiltonian for the duplicated system is given by
$$ H(\underline s) = -\sum_{\langle i,j \rangle}[\Cal E^0_{i,j}\delta_{s_i
= s_j} + \Cal E^1_{i,j}\delta_{s_i = s_j+1} +  \Cal E^1_{i,j}\delta_{s_i =
s_j-1}]
\tag 5
$$ where the $\delta$'s are one if the indicated condition is satisfied and
vanish otherwise and we have used additive notation for the group structure
of the spin--states.  Explicitly, $s\to s+1$ means
$(++)\to (+-) \to (--) \to (-+) \to (++)$.
\footnote {We remark that the Hamiltonian in Eqn\. (5) differs from (two
copies of) that in Eqn\. (5)) -- with $h_i \equiv 0$ -- by an overall
constant}

For each $\langle i,j \rangle$ we may expand the statistical weight
associated with $H$ as:
$$ e^{\beta\Cal E^0_{i,j}\delta_{s_i = s_j}} e^{\beta\Cal
E^1_{i,j}\delta_{s_i = s_j+1}} e^{\beta\Cal E^1_{i,j}\delta_{s_i = s_j-1}}
= 1 + B_{i,j}\delta_{s_i = s_j} + R_{i,j}\delta_{s_i = s_j+1} +
R_{i,j}\delta_{s_i = s_j-1}
\tag 6
$$ with $B_{i,j} = e^{\beta\Cal E^0_{i,j}} - 1$ and
$R_{i,j} = e^{\beta\Cal E^1_{i,j}} - 1$. Note that no terms involving
products of
$\delta$'s appear since the conditions are exclusive. Thus, there are  four
possibilities for each $\langle i,j \rangle$: A ``vacant'' bond
corresponding to the 1, a ``blue'' bond corresponding to the term
$B_{i,j}\delta_{s_i = s_j}$  and one of two types of ``red'' bonds which
forces $s_i = s_j \pm 1$.

A graphical configuration $\underline \eta$ is obtained by a selection of
one of the above four terms for each bond of the lattice.  The weight of
$\underline \eta$ is given by
$$
\underline V(\underline \eta) =
\sum_{\underline s}\prod_{\langle i,j \rangle}
\Cal R_{\langle i,j \rangle}(\underline \eta)
\delta_{\langle i,j \rangle}(\underline \eta, \underline s)
\tag 7
$$ where $\Cal R_{\langle i,j \rangle}(\underline \eta) = 1$, $B_{i,j}$ or
$R_{i,j}$ according to whether a vacant, blue or red bond has been selected
and $\delta_{\langle i,j \rangle}(\underline \eta, \underline s)$ is the
corresponding delta factor (or 1 for the vacant bonds).

These weights are not particularly  difficult to evaluate explicitly nor
are the results particularly illuminating.  Of greater interest is the
weight associated with all configurations $\underline \eta$ with the same
colored bonds but different types of {\it red} bonds.  We will denote these
equivalence classes by an unadorned $\eta$ and define
$V(\eta) = \sum_{\underline \eta \in \eta}\underline V(\underline \eta)$.
An explicit formula for $V(\eta)$ is also possible which we will give after
the relevant notation is developed.

First it is noted that not all $\eta$'s have non--zero weight.  In
particular, it is necessary and sufficient that each closed loop of
non--vacant (hereafter ``occupied'') bonds of $\eta$ must consist of an
even number of red bonds.  Let $D(\eta)$ be the indicator that the above
condition is satisfied.  Next, let $F(\eta)$ denote the ``bond fugacity''
prefactor (which is the same for all $\underline \eta$'s in the equivalence
class)
$$ F(\eta) = \prod_{\langle i,j \rangle}
\Cal R_{\langle i,j \rangle}(\eta).
\tag 8
$$ Finally, the geometric factors:  It is observed that there are two
relevant notions of connectedness in a configuration $\eta$, namely
connected by blue bonds or connected by red and blue bonds, which we will
call a {\it grey} connection.  Let
$C_b(\eta)$ denote the number of blue components in the configuration
$\eta$ and
$K(\eta)$ the number of components by the less stringent (grey)
definition.  Thus
$K(\eta) \leq C_b(\eta)$; for example, if each bond is occupied and red,
$K(\eta) = 1$ while
$C_b(\eta)$ is the total number of sites.

With the above in hand, the result is
$$ V(\eta) = F(\eta)D(\eta)2^{K(\eta)}2^{C_b(\eta)}.
\tag 9
$$ We remark that we have not yet discussed boundary conditions; the above
is strictly true only for free boundary conditions on an arbitrary (but
finite) graph. We denote the associated measure by $v(-)$ with various
super-- and sub--scripted adornments to be appended as needed to indicate
the lattice, the boundary conditions and the fields.

We introduce another measure, called the Edwards--Sokal measure, for the
representation.  This is a joint measure on bond and spin configurations
that assigns an equal weight to each spin configuration that is
``consistent'' with the given bond configuration:
$$ V^{ES}(\underline s, \eta) = F(\eta)\Delta(\underline s, \eta).
\tag 10
$$ In the above, $\Delta(\underline s, \eta)$ vanishes unless (a) the spin
values are constant in each blue component of $\eta$ and (b) the spin values
$s_i$ and $s_j$ across each red bond of $\eta$ differ by one:
$s_i = s_j \pm 1$.  If (a) and (b) are both satisfied then
$\Delta(\underline s, \eta) = 1$.  The Edwards--Sokal measures defined by
these weights will be denoted by $v^{ES}(-)$.  It is not difficult to see
that
$$
\sum_\eta V^{ES}(\underline s, \eta) = e^{-\beta H(\underline s)}
\tag 11a
$$ and
$$
\sum_{\underline s}V^{ES}(\underline s, \eta) = V(\eta).
\tag 11b
$$ Thus the bond marginal of $v^{ES}(-)$ is the red/blue random cluster
measure and the spin marginal is the Gibbs measure for the system. The
Edwards--Sokal measure facilitates an interpretation of the bond
configuration
$\eta$.  In particular, connected clusters of blue sites represent spins
that are all in the same state, separate clusters not connected by any
bonds are uncorrelated from one another and sites connected by an even
(odd) number of red bonds represent spins that agree mod 2 (differ by
$\pm 1$).

Boundary conditions are readily accounted for in the red/blue random
cluster measures.  If $\Lambda$ is the graph and $\partial \Lambda$ the
boundary sites, a single spin specification on $\partial \Lambda$ divides
$\partial \Lambda$ into (at most) four components.  These components are
treated as a
single site in the counting of $K(\eta)$ and
$C_b(\eta)$.   Furthermore, there are restrictions on the configuration
$\eta$, e\.g\. no two of the boundary components can be connected by blue
bonds.  Of particular interest are the wired boundary conditions where all
the boundary sites are treated as a single component.  In the absence of
fields, the boundary type may be identified as any of the four states or
superpositions thereof.

We now describe a cluster algorithm based on the red/blue representation.
Starting from a spin configuration $\underline s$, each
$\langle i,j \rangle$ is checked for the conditions $s_i = s_j$ or $s_i =
s_j \pm 1$.  If
$s_i = s_j$, a blue bond is placed on
$\langle i,j \rangle$ with probability $b_{i,j} = B_{i,j}/[1 + B_{i,j}]=1 -
e^{-4\beta J_{i,j}}$ or is declared vacant with probability $1 - b_{i,j}$.
If
$s_i = s_j \pm 1$, a red bond is placed on
$\langle i,j \rangle$ with probability $r_{i,j} = R_{i,j}/[1 + R_{i,j}]=1 -
e^{-2\beta J_{i,j}}$ or the bond is declared vacant with probability $1 -
r_{i,j}$.  On all bonds for which $s_i$ and
$s_j$ are completely opposed ($s_i = s_j + 2$, i\.e\. $s_i = -s_j$) the
bond is always vacant.  All bond moves are performed independently.  This
creates the configuration
$\eta$.  Given the configuration $\eta$, the updated spin configuration is
created by two  types of cluster moves.  First, the spins in each isolated
grey cluster can be coherently rotated by 0, $\pm 1$, or 2 with equal
probability. Second,  the spin type of all spins in a blue cluster
(including single spins with no blue bonds) can be reversed ($s \to -s$)
with probability 1/2. We will refer to these as grey moves and blue moves
respectively.  The cluster moves are all performed independently. This
algorithm is of the Swendsen--Wang \cite{SW} type and it is straightforward
to show, following \cite{ES}, that it simulates the Edward--Sokal measure
of Eqn\. (10).  For more details (and more formal descriptions) concerning
general algorithms of this sort, the reader should consult
\cite{CM I}, \cite{CM II}

\remark{Remark} The strength of this representation is that percolation
(essentially of blue bonds) can be directly related to the
non--vanishing of the spontaneous magnetization.  This will be discussed in
full detail in the next section.  One principal difficulty is the presence
of large grey clusters which, under a variety of conditions, will hinder
rapid equilibration.  Indeed, at moderate temperatures above the magnetic
ordering temperature, there is, generically, percolation of grey bonds.
(For example, this can be established by rigorous bounds for the $2d$
square lattice.)  Furthermore at low temperatures, the rapid ``tunneling''
events characteristic of better representations are inhibited. For example
consider a situation where $\beta \gg 1$ and suppose that all the spins on
the left side of a box are in the state $++$ while on the right side of the
box all spins are $+-$.  Here both the left and right sides will be tied up
in a single cluster.  If we use two independent SW algorithms (for the
Ising case, $\Cal E^0 = 2\Cal E^1$ or, in the general Ashkin--Teller case
the bilayer representation) the two clusters are disjoint and the interface
will disappear in a few Monte Carlo steps.  However with the red/blue
representation, there would nearly always be  red bonds connecting the left
and right regions which prevent  full alignment of the box; in this case,
an extremely long time is required for equilibration.  This represents the
principal reason that this approach was abandoned as a tool to study the
Ashkin--Teller model.  However, as we shall see, in the presence of
external fields, the red/blue representation has significant advantages
over the alternatives.
\endremark
\subhead Red/Blue Representation with Fields
\endsubhead Before we actually introduce the fields, let us present one
more refinement which, in the absence of fields is somewhat redundant.
Consider a configuration
$\eta$. Recall that each blue cluster of $\eta$ represents spins that are
all in the same state.  Now a blue cluster may be part of some larger grey
cluster but the dynamics described in the previous section allows
the spins of each blue cluster to flip ($++ \leftrightarrow --$, $+-
\leftrightarrow -+$) leaving the remaining spins of the grey cluster
invariant.  But then a further classification is allowed for the blue
clusters namely whether the cluster is in the $(++,--)$ or $(-+,+-)$
states.  In the former case, we will say that the cluster is {\it
dark--blue} and in the latter, {\it light--blue}.  As we shall see,
percolation of the light--blue bonds represents the onset of magnetic
ordering.

We will denote these red, dark--blue and light--blue (RDLB) configurations
by
$\Xi$'s. It is seen that the weight of a configuration $\Xi$ is given by
$$
\bold V(\Xi) = F(\Xi)\Bbb D(\Xi) 2^{C_b(\Xi)}
\tag 12
$$ where $F(\Xi)$ has  its previous meaning, $C_b(\Xi)$ is the number of
blue clusters -- both dark and light -- and $\Bbb D(\Xi)$ indicates that
within each grey cluster, all paths between pairs of light--blue clusters
consist of an even number of red bonds, similarly for pairs of dark--blue
clusters while any path between a dark--blue and a light--blue cluster
consists of an odd number of red bonds.  We note that the ``grey moves'' of
the algorithm described in the previous section (where the spins of the
grey clusters were coherently rotated) consists, with probability 1/2, of
exchanging dark--blue for light--blue within each grey cluster.  As such it
is easy to recover the weights in Eqn\. (9) from those of Eqn\. (12) -- for
each grey cluster, one sums over both these possibilities.  We denote the
measure defined by the weights in Eqn\. (12) by
$\Bbb V(-)$ and the associated Edwards--Sokal measure by $\Bbb V^{ES}(-)$.

Finally we remark that the division of blue into dark and light degrees of
freedom creates more candidates for the boundary conditions.  In particular
we must now separately consider {\it light--wired} boundary conditions
corresponding, in the spin system, to $+-$ or $-+$ boundary conditions or
{\it dark--wired} corresponding to $++$ or
$--$ boundary conditions.  We reiterate that in the absence of external
fields, these are all equivalent:  the graphical problem is invariant under
the exchange of dark and light labels.

We are finally ready to introduce the fields.  Consider the Ising
Hamiltonian as given in Eqn\. (1).  Upon duplication we have
$$
H = H(\underline s) = \Cal H(\underline \sigma) + \Cal H(\underline
\tau) =-
\sum_{\langle i,j \rangle}J_{i,j}[\sigma_i\sigma_j + \tau_i\tau_j] -
\sum_ih_i[\sigma_i + \tau_i].
\tag 13
$$ The result is clear:  the fields do not act on the
spins $s_i$ that are in the states ($+-$) or ($-+$)  and hence on the
light--blue clusters.  In the construction of the graphical representation,
we may either go the route of ghost bonds (as in Eqn\. (4)) or  field
energies (as in Eqn\. (5)).  The latter choice is somewhat easier and
gives us the weights
$$
\bold{V}(\Xi) = F(\Xi)\Bbb D(\Xi) 2^{C_{lb}(\Xi)}\prod_{\alpha =
1}^{C_{db}(\Xi)} 2 \cosh[ 2 \bold h(c^{db}_\alpha)]
\tag 14
$$ where $c^{db}_\alpha$ is the $\alpha^{\text{th}}$ dark--blue cluster of
$\Xi$ and
$\bold h(c^{db}_\alpha)$ is defined by
$$
\bold h(c^{db}_\alpha) = \beta\sum_{i \in c^{db}_\alpha}h_i.
\tag 15
$$
We again emphasize that these weights are non-negative and therefore
define a probability measure on the space of all RDLB configurations.  As
above the measures will be denoted by a $\Bbb V$; in particular for the
Hamiltonian $H$ at inverse temperature $\beta$, in volume $\Lambda$ with
boundary condition $*$ we will denote
these measures by $\Bbb V_{\Lambda^*;\beta, H}$.

A cluster algorithm based on the RDLB representation  is similar to the
algorithm described in the previous section.  The bond moves are
identical.   There are now three kinds of spin moves; grey, dark--blue and
light--blue. The dark--blue cluster flips occur with probabilities weighted
by the exponent of their field energies:  For the cluster
$c^{db}_\alpha$ the ratio of forward to backward rates is $e^{\pm2\bold
h(c^{db}_\alpha)}$ with the sign depending on the initial and final spin
state. The grey cluster move exchanges light and dark blue cluster with a
probability that depends on the fields.  Although the grey and dark--blue
moves depend on the fields in a complicated way, the light-blue move is
independent of the field. The crucial point is that since the field energy
is always zero for the light--blue clusters, these are ``freely'' flipped
with probability 1/2.  As we demonstrate in the next section, for all
ferromagnetic Ising problems, the light--blue infinite cluster density can
be identified with the magnetic order parameter.   Thus, in this
representation, the degrees of freedom contributing to the long range order
are decoupled from the fields. Nevertheless, the algorithm may experience
slowing due to the presence of large grey or dark--blue clusters.  In the
final section, we will discuss some strategies that can be used to overcome
these shortcomings.

\subhead The order parameter
\endsubhead In the following we present some elementary results for Ising
systems. Although most of these are proved explicitly or implicitly in the
standard treatments (e\.g\. \cite{G}) for completeness we will provide a
proof sketch. Below we will assume that $\Lambda$ is a finite graph that is
a subset of some infinite $\Bbb L$ and the Ising Hamiltonian, $\Cal H$ is
as given in Eqn\. (1).  We will consider sequences $(\Lambda_k)$ which
``tend to $\Bbb L$'' meaning $\Lambda_{k+1} \supset \Lambda_k$ and each
$i\in \Bbb L$ is, eventually, in some $\Lambda_k$.  This is denoted by
$\Lambda \nearrow \Bbb L$.  Thermal states on $\Lambda$ at inverse
temperature $\beta$ and boundary condition
$\#$ on $\Bbb L \setminus \Lambda$ will be denoted by
$\langle - \rangle_{\Lambda;\beta,\Cal H}^\#$.  The special (and most
important) cases are
$\# = +$ ($\# = -$) corresponding to each spin of $\Bbb L
\setminus \Lambda$ set to $+$ ($-$).
\proclaim{Proposition 1} Let $i \in \Bbb L$ and for each $\Lambda$ with $i
\in \Lambda$ let
$$
m_i^+(\Lambda) = \langle \sigma_i \rangle_{\Lambda;\beta,\Cal H}^+
$$
and
$$
m_i^-(\Lambda) = \langle \sigma_i \rangle_{\Lambda;\beta,\Cal H}^-
$$
Then the limits
$$
m_i^+ = \lim_{\Lambda \nearrow \Bbb L}m_i^+(\Lambda)
$$
and
$$
m_i^- = \lim_{\Lambda \nearrow \Bbb L}m_i^-(\Lambda)
$$ exist for any sequence $(\Lambda_k)$ as described above and are
independent of the sequence.  Let $\mu_i(\beta) = \frac 12[m_i^+(\beta) -
m_i^-(\beta)]$ denote the ``magnetization excess''.  Then the necessary and
sufficient condition for Gibbsian uniqueness is that for each $i$,
$\mu_i(\beta) = 0$.
\endproclaim
\demo{Proof} We use the fact that the finite volume measures (and hence the
infinite limits thereof) have the strong FKG property for the ordering
induced by
$(\sigma_i = +) > (\sigma_i = -)$.  This means that the $+$ boundary
conditions on
$\Lambda$ are ``higher'' than any other boundary conditions.  In
particular, if
$\Lambda^\prime \supset \Lambda$ the restriction of
$\langle - \rangle_{\Lambda^\prime;\beta,\Cal H}^+$ to $\Lambda$ is
dominated by
$\langle - \rangle_{\Lambda;\beta,\Cal H}^+$.  This ensures the existence
of an unambiguous limiting plus state (independent of the sequence
$(\Lambda_k)$) which will be denoted by $\langle -
\rangle^+_{\beta,\Cal H}$ as well as the existence of the $m_i^+$.
Similar considerations provide the infinite volume state
$\langle - \rangle^-_{\beta,\Cal H}$ and  the
$m_i^- \equiv \langle \sigma_i \rangle^-_{\beta,\Cal H}$. If, for
any $i$,
$2\mu_i = m_i^+ - m_i^-$ is not zero then obviously there is more than one
state.  On the other hand if for every $i$, $\mu_i = 0$ then, by
Strassen's theorem, \cite {S} we have
$\langle - \rangle^+_{\beta,\Cal H} = \langle -
\rangle^-_{\beta,\Cal H}$. But since any limiting state lies above
$\langle - \rangle^-_{\beta,\Cal H}$ and below
$\langle - \rangle^+_{\beta,\Cal H}$ (in the sense of FKG) it
follows that the limiting state is unique.
\qed
\enddemo
\remark{Remark} Under the usual physical assumptions that  $\Bbb L = \Bbb
Z^d$ or some other regular $d$--dimensional lattice and that the $h_i$ and
$J_{i,j}$ are periodic or i.i.d. random variables then
$$
\mu(\beta) = \lim_{\Lambda \nearrow \Bbb L}\frac{1}{|\Lambda|}
\sum_{i \in\Lambda}\mu_i(\Lambda) \equiv
\lim_{\Lambda \nearrow \Bbb L}\frac{1}{|\Lambda|}
\sum_{i \in\Lambda} \frac 12[m_i^+(\Lambda) - m_i^-(\Lambda)]
$$ exists (almost surely in the random cases) for all $\Lambda \nearrow
\Bbb L$ in the sense of Van Hove.  The quantity $\mu(\beta)$ may be
identified with a thermodynamic derivative.  Indeed if $f(\goth b)$ is the
free energy in the presence of the applied field $\goth b$ (that is to say
the term $-\sum_i\goth b\sigma_i$ is added to $\Cal H$) then
$\mu(\beta) = -\frac 12\beta[f^\prime(\goth b = 0^+) - f^\prime(\goth b =
0^-)]$.
\endremark Our main result:
\proclaim{Theorem 2} Let $\Lambda$ denote a finite graph (which may be
regarded as a subset of some infinite $\Bbb L$) and consider the
light--blue wired measure
$\Bbb V _{\Lambda^w;\beta, H}(-)$ on RDLB configurations as described in
Eqn\. (13)--(15) with wired boundary conditions.  Let $i\in\Lambda$ and let
$\{i\leftrightarrow \partial \Lambda\}$ denote the event that $i$ is
connected to the boundary by light--blue bonds.  Then
$$
\Bbb V _{\Lambda^w;\beta, H}(\{i\leftrightarrow \partial \Lambda\}) = \frac
12[m_i^+(\beta,\Lambda) - m_i^-(\beta,\Lambda)]
\equiv
\mu_i(\beta;\Lambda).
$$
\endproclaim
\demo{Proof} Let $\langle -\rangle_{\Lambda;\beta,H}^{+-}$ denote the
thermal state for the (four--state spin) model on $\Lambda$ with
Hamiltonian $H$ as in Eqn\. (13) and $+-$ boundary conditions. It is clear
that if $A$ is any function of the $\sigma$'s alone then
$\langle A \rangle_{\Lambda;\beta,H}^{+-} = \langle A
\rangle_{\Lambda;\beta,\Cal H}^+$ (because the $\sigma$'s and the $\tau$'s
are independent) and similarly if $A^\prime$ is a function of the $\tau$'s
alone then
$\langle A^\prime \rangle_{\Lambda;\beta,H}^{+-} = \langle A^\prime
\rangle_{\Lambda;\beta,\Cal H}^-$.
 Thus
$$
\frac 12\langle [\sigma_i - \tau_i] \rangle_{\Lambda;\beta,H}^{+-} =
\frac 12 [m_i^+(\Lambda) - m_i^-(\Lambda)] = \mu_i(\Lambda).
\tag 16
$$ The stated identity follows easily from the Edwards--Sokal measure.  We
start with
$$
\frac 12 \langle [\sigma_i - \tau_i] \rangle_{\Lambda;\beta,H}^{+-} =
\Bbb E^{ES}_{\Lambda^{lw}}(\frac 12[\sigma_i - \tau_i])
\tag 17
$$ where $\Bbb E^{ES}_{\Lambda^{lw}}(-)$ denotes expectation with respect
to the measure
$\Bbb V^{ES}_{\Lambda^{lw}}(-)$ with light--blue wired boundary conditions
on
$\partial \Lambda$.  Let us consider the three possibilities for the site
$i$: (1) the event $\{i\leftrightarrow \partial \Lambda\}$ occurs, (2) the
site $i$ belongs to a dark--blue cluster and (3) the site $i$ belongs to a
light--blue cluster that is not connected by light--blue bonds to the
boundary.  We may express the right hand side of Eqn\. (17) in terms of the
three associated conditional expectations.

Whenever the third possibility occurs, regardless of whether $i$ is grey
connected to
$\partial \Lambda$,  the cluster of $i$ is $+-$ and $-+$ each with
probability 1/2.  If
$i$ is in a dark--blue cluster then $\sigma_i = \tau_i$.  Thus, items (2)
and (3) contribute nothing.  On the other hand, if $\{i\leftrightarrow
\partial \Lambda\}$ occurs then the boundary conditions force
$\sigma_i = +$ and  $\tau_i =
-$. This establishes the stated identity.
\qed
\enddemo

\remark{Remark} It is observed that this representation may also be used
for the calculation of correlation functions.  For example, consider an
Ising system in finite volume $\Lambda$ with fields $(h_i)$ and boundary
condition $*$.   Duplicating the system with the {\it same} boundary
conditions it is not hard to see that the expectation of $(\sigma_i -
\tau_i)(\sigma_j - \tau_j)$ is exactly twice the truncated spin--spin
correlation function between the sites $i$ and $j$.  Now if the site $i$
belongs to a dark--blue cluster -- which is necessitated if it is connected
to the boundary -- then the contribution is zero because $\sigma_i =
\tau_i$.  Similarly if $j$ belongs to a  dark--blue cluster.  Furthermore,
if $i$ and
$j$ belong to separate light--blue clusters there is cancellation because
for fixed $\tau_j$ and
$\sigma_j$, the quantity $(\sigma_i - \tau_i)$ takes on the value 2 and
$-2$ with equal probability.  Finally, if $i$ and $j$ are in the same
light--blue cluster, the result is four.  Thus we are left with another
identity:  The truncated spin--spin correlation function is half the
probability that the relevant sites are in the same light blue cluster.
\endremark

With the exception of a few technical points the picture is complete.
Given $+-$ (i.e. light--blue wired) boundary conditions, there either is or
isn't percolation of light--blue bonds.  If there is, the stated order
parameter is positive and if not, there is no ordering of the stated type.
In fact, these results apply to other systems of this type.   For example,
the previous sentence holds generally for the  Ashkin--Teller model in an
external field where $0 \leq \Cal E^1_{i,j} \leq \frac 12\Cal E^0_{i,j}$.
However, specific to the Ising system is the stronger statement that if the
magnetic excess vanishes ($\mu(\beta) \equiv 0$), there is no long--range
order of {\it any} type.  There is a second issue that is not unrelated to
the first.  We have implicitly ``defined'' percolation to mean percolation
in the state that ultimately produces the order parameter.  (Of course, in
simulations these are rarely the boundary conditions that are actually
used.)  What has not been ruled out is the possibility that there is
light--blue percolation in {\it some} limiting state but not in the
limiting $+-$ state.  Although there can be little doubt that in general
this does not occur, for Ising systems we have a complete proof.

We start with a precise definition of (light--blue) percolation:
\definition{Definition} Let $\Lambda \subset \Bbb L$ be a finite lattice
and consider the various measures
$\Bbb V _{\Lambda^\#;\beta}(-)$ on RDLB configurations with boundary
conditions
$\#$.  For $i\in\Lambda$, consider the probability that $i$ is connected to
$\partial
\Lambda$ by light--blue bonds maximized over all possible boundary
conditions:
$$ P_i(\beta;\Lambda) = \underset{\#} \to{\text{max}}
\Bbb V _{\Lambda^\#;\beta,H}(\{i \leftrightarrow \partial
\Lambda\}).
\tag 18a
$$ We define
$$ P_i(\beta;\infty) = \lim_{\Lambda \nearrow \Bbb L} P_i(\beta;\Lambda).
\tag 18b
$$ (It is not hard to show that this limit exists independent of the
sequence $(\Lambda_k)$.)  If $P_i(\beta;\infty) \neq 0$ for any $i$, we say
there is
percolation and if $P_i(\beta;\infty) \equiv 0$ there is no percolation.
\enddefinition For Ising systems, we have the following characterization:
\proclaim{Theorem 3} For an Ising system as defined by the Hamiltonian in
Eqn\. (1) there is a unique limiting state if and only if there is no
percolation.
\endproclaim
\demo{Proof} By Proposition 1, there is uniqueness iff $\mu_i = 0$ for all
$i$. Obviously
$\mu_i(\beta) \leq P_i(\beta;\infty)$.  It remains to be established that
$\mu_i(\beta) = 0 \Longrightarrow P_i(\beta;\infty) = 0$.  Here an argument
similar to that used in
\cite {CM II} Theorem 3.8 may be applied.  Let $*$ denote any boundary
conditions that optimize $\Bbb V _{\Lambda^*;\beta}(\{i \leftrightarrow
\partial
\Lambda\})$ which, without loss of generality, is taken to be a single--spin
specification. There are (at most) four components to $\partial \Lambda$:
$\partial \Lambda^{++}(*)$,
$\partial
\Lambda^{--}(*)$ and $\partial \Lambda^{\pm\mp}(*)$ corresponding to the four
spin--types.
We may write $P_i(\beta;\Lambda) =
P_i^{+-}(\beta;\Lambda) + P_i^{-+}(\beta;\Lambda)$ with
$$
P_i^{+-}(\beta;\Lambda) =
\Bbb V _{\Lambda^*;\beta,H}(\{i \leftrightarrow \partial \Lambda^{+-}(*)\})
\tag 19
$$
etc.  By $\underline \sigma \leftrightarrow \underline \tau$ symmetry,
it may be assumed that $P_i^{+-}(\beta;\Lambda) \geq
P_i^{-+}(\beta;\Lambda)$.

We claim that in the light--blue wired (i.e. $+-$) boundary conditions, the
probability of
$\{i \leftrightarrow \partial \Lambda^{+-}(*)\}$
is at least as large as it is in the
$*$ boundary conditions and hence, after the infinite volume limit,
$\mu_i(\beta) \geq
\frac 12P_i(\beta; \infty)$.  Let us consider the situation from the
perspective of the Edwards--Sokal measure.  Decomposing
$P_i^{+-}(\beta;\Lambda)$ according to spin configurations we have
$$
\Bbb V _{\Lambda^*;\beta,H}
(\{i \leftrightarrow \partial \Lambda^{+-}(*)\}) =
\sum_{\underline \sigma}
\Bbb V^{ES}_{\Lambda^*;\beta,H}(\underline \sigma,\underline \tau)
\Bbb V^{ES}_{\Lambda^*;\beta,H}(\{i \leftrightarrow \partial
\Lambda^{+-}(*)\}\mid\underline \sigma,\underline \tau)
\tag 20a
$$
The right hand side is of the form of a thermal expectation of a
function of spin configurations, namely
$\Bbb V^{ES}_{\Lambda^*;\beta,H} (\{i \leftrightarrow \partial
\Lambda^{++}(*)\}\mid\underline \sigma,\underline \tau)$;
$$
\Bbb V _{\Lambda^*;\beta,H}(\{i \leftrightarrow \partial \Lambda^{+-}(*)\})
= \langle  \Bbb V^{ES}_{\Lambda^*;\beta,H} (\{i \leftrightarrow \partial
\Lambda^{+-}(*)\}\mid\underline
\sigma,\underline \tau)\rangle_{\beta,H;\Lambda}^{*} .
\tag 20b
$$ However, in light of the algorithmic construction of the configurations
$\Xi$ given $\underline
\sigma$ it is clear that this function is increasing in $\underline
\sigma$ and decreasing in $\underline \tau$:  Indeed, this is amounts to a
bond--site percolation problem and increasing the number of plus
$\sigma$'s and minus $\tau$'s just increases the number of sites. But then,
by the FKG property of the Gibbs measures,
$$
\align
\langle  \Bbb V^{ES}_{\Lambda^*;\beta,H} (\{i \leftrightarrow \partial
\Lambda^{+-}(*)\}\mid\underline
\sigma, \underline \tau)\rangle_{\Lambda;\beta,H}^{*} &\leq
\langle  \Bbb V^{ES}_{\Lambda^w;\beta,H} (\{i \leftrightarrow \partial
\Lambda^{+-}(*)\}\mid\underline
\sigma,\underline \tau)\rangle_{\Lambda;\beta,H}^{+-}\\ &=
\Bbb V _{\Lambda^w;\beta,H}(\{i \leftrightarrow \partial \Lambda^{+-}(*)\})
\leq \mu_i(\beta;\Lambda)
\tag 21
\endalign
$$ from which the stated result follows.
\qed
\enddemo
\subhead Conclusions
\endsubhead The upshot of the previous analysis is that the crucial degrees
of freedom at points where there is a magnetic ordering transition are
represented by the light--blue clusters.  For the algorithms described, the
spin--transitions within these clusters is uninhibited and this should lead
to substantial improvement over other cluster methods in many situations.
On the other hand, when the system is critical or at a point of phase
coexistence, it may be too much to hope that there are never any other
large--scale fluctuations.  In particular, in the absence of fields, by
light/dark symmetry, the distribution of dark--blue clusters is identical
to that of the light--blue.  In the presence of fields, there might still
be large dark--blue clusters and these would have a tendency to
``freeze''.  Furthermore, since we expect grey percolation at all points of
physical interest, the transitions that exchange dark with light could be
disastrously slow.

However even assuming the worst case scenario, namely that the only aspects
worth preserving are the light--blue moves, the algorithm may be
supplemented with Monte Carlo steps from other algorithms to insure
ergodicity.  In these circumstances, the whole of the previously described
algorithm need not be implemented.  Indeed, the following ``pure
light--blue'' move clearly satisfies detailed balance:  Starting from a
spin--configuration
$\underline s$, between all neighboring pairs $\langle i,j \rangle$ where
$s_i = s_j = +-$ or $s_i = s_j = -+$ independently place a light--blue bond
with probability $1 - e^{-4\beta J_{i,j}}$.  Clusters are identified and,
independently, each cluster is flipped or left alone with probability
$\frac 12$.  Such moves (which are obviously not ergodic) can then be
sandwiched between single--spin updates.

In addition (for Ising systems) let us recall that it is only the
statistical behavior of single layers that is actually of interest.  Thus
we may consider an ensemble of replicas where, in the cluster moves,
members of the ensemble are paired randomly.  Furthermore, if there is some
translation invariance (e\.g\. the staggered--field problem) then the
replicas may be translated relative to one another. Of course neither of
the above moves will restore ergodicity; other types of updates  are
still required.

A closely related set of ideas described in
\cite{HBb} and \cite{HBc} has recently been brought to our attention. Here
the idea is to ``fold'' the lattice through an axis of symmetry; e\.g\. the
$x_1$--axis.  Spins at the sites
$i = (x_1, x_2, \dots x_d)$ and $i^* = (-x_1, x_2, \dots x_d)$ are now
regarded as ``pairs'' Assuming that $h_i = h_{i^*}$ (as would be the case
for the staggered field problem) one may implement a procedure to generate
light--blue clusters as described above.
In the references cited, single cluster methods were used.  Aside from the
``boundary spins'', i\.e\. the $x_1$--axis, the method of folding and of
duplication are, for all intents and purposes, identical for problems with
periodically varying fields.  However, the advantage of having two separate
replicas is manifestly apparent for systems where $h_i$ is random or, more
generally, has no spatial periodicity.

Recently, the authors tested some of these strategies  on the square lattice
staggered--field problem \cite{RMC}.   The algorithm was applied to a
two-replica system and consisted of light--blue cluster moves, random
translations of one replica relative to the other by even or odd lattice
vectors, and Metropolis updates.  In the case of odd translations, one
replica is also globally flipped relative to the other.  In \cite{RMC} we
use the terminology ``active sites'' (``inactive sites'') to refer to sites
in light--blue (dark--blue) clusters since, in the cluster part of the
algorithm, only active sites are flipped. The results were quite
satisfactory; for systems up to scale 256, our data is consistent with a
critical slowing exponent less than $0.5$  (compared with local dynamics
where the value is found to be in excess of 2).

\enddocument
\end